\documentclass{article}

     \PassOptionsToPackage{numbers, compress}{natbib}

     \usepackage[final]{neurips_2024}

\usepackage[utf8]{inputenc} 
\usepackage[T1]{fontenc}    
\usepackage{hyperref}       
\usepackage{url}            
\usepackage{booktabs}       
\usepackage{amsfonts}       
\usepackage{nicefrac}       
\usepackage{microtype}      
\usepackage{xcolor}         

\title{Rules, Cases, and Reasoning: Positivist Legal Theory as a Framework for Pluralistic AI Alignment}

\author{%
  Nicholas A. Caputo\thanks{I am grateful to Mackenzie Arnold, Noah Feldman, Ioannis Kalpouzos, Lawrence Lessig, Cass Sunstein, Justin Walker, Zachary Wojtowicz, and Jonathan Zittrain for their insights, comments, and help.} \\
  Oxford Martin AI Governance Initiative \\
  Oxford, UK OX1 3BD \\
  \texttt{ncaputo@jd24.law.harvard.edu} \\
}

\begin{document}

\maketitle

\begin{abstract}
  Legal theory can address two related key problems of alignment: pluralism and specification. Alignment researchers must determine how to specify what is concretely meant by vague principles like helpfulness and fairness and they must ensure that their techniques do not exclude alternative perspectives on life and values. The law faces these same problems. Leading legal theories suggest the law solves these problems through the interaction of rules and cases, where general rules promulgated by a democratic authority are given specific content through their application over time. Concrete applications allow for convergence on practical meaning while preserving space for disagreement on values. These approaches suggest improvements to existing democratic alignment processes that use AI to create cases that give content to rules, promoting more pluralist alignment.
\end{abstract}

\section{Introduction}

The law must confront questions of how to reconcile competing values and views of what society is and should be by nature of its very form. It is a set of general rules, made by bodies recognized as having lawmaking authority, that bind all members of a society in the specificity of their daily conduct despite fundamental disagreements on how to live and what matters \cite{Hart}. These general rules are applied to disputes in cases that, in a system of precedent, bind others in future disputes on similar points of law \citep{Goodhart, Deutsch, Schauer}. How is it possible to preserve a pluralist society that respects the free choice of individuals to make their own decisions given the necessary extent to which general rules reduce individual freedom of choice? And how can general rules predictably be given concrete meaning given the ambiguity of language and changes in circumstances? 

Alignment, particularly the kind of finetuning alignment that underlies leading large language models, \citep{askell2021generallanguageassistantlaboratory, bowman2022measuringprogressscalableoversight, kenton2021alignmentlanguageagents, kirk2023pastpresentbetterfuture, perez2022discoveringlanguagemodelbehaviors, rafailov2024directpreferenceoptimizationlanguage, korbak2023pretraininglanguagemodelshuman, scheurer2022traininglanguagemodelslanguage} faces these same problems. Any principle or rule put into a model shapes its outputs in a way that others, including users, might find incompatible with their values, and researchers are still not certain how models interpret or will apply the rules put into them \citep{Gabriel, hendrycks2023aligningaisharedhuman, Templeton, kasirzadeh2022conversationartificialintelligencealigning, ganguli2023capacitymoralselfcorrectionlarge}. As Kundu et al. \cite{kundu2023specificversusgeneralprinciples} note, simply giving a rule to a model means that the model will interpret the rule based on its underlying black-boxed sense of the words in which the rule is expressed rather than according to what the researchers wanted. Furthermore, there may be significant tradeoffs between alignment to user preferences and to broader social values \cite{kirk2023personalisationboundsrisktaxonomy, chen2023largelanguagemodelsmeet}. Leading approaches are either the product of small groups in AI labs or, where democratic inputs have been sought, of majorities of those few selected to provide such input \citep{christiano2023deepreinforcementlearninghuman, bai2022constitutionalaiharmlessnessai, GanguliCCAI, CIP}. Neither approach promotes pluralism. Pluralistic models must instead be able to reason contextually and from different perspectives, drawing on the body of alignment rules and on normal linguistic practice to ensure that there is consistency and predictability in their outputs but also taking the perspective of minorities with different views from those predominant in training data or in labs \citep{bakker2022finetuninglanguagemodelsagreement, kirk2024prismalignmentprojectparticipatory, sorensen2024roadmappluralisticalignment, Sorensen_2024}.

Legal theory provides ideas that might aid in solving these problems. The law specifies its general rules through the application of those rules in concrete cases. In each factual dispute that forms a case, an interpretation of the rule is tested and becomes precedent \cite{Barrett}. Over time, these interpretations accumulate and map out not just the core meaning of a rule but also how to think about its edge cases \cite{Hart}. Reasoning about concrete facts rather than high level justifications allows pluralism to flourish because members of society can coordinate on concrete ways to live while respecting each other’s values \citep{Sunstein1, Sunstein2}. Alignment researchers have begun to draw on legal concepts to align their systems \citep{bai2022constitutionalaiharmlessnessai, Claude_Const, GanguliCCAI, feng2023caserepositoriescasebasedreasoning, OAI_democ_inputs, bignotti2024legalmindsalgorithmicdecisions, Gordon_2022, irving2018aisafetydebate, browncohen2023scalableaisafetydoublyefficient}. Incorporating the interplay between rules and cases at the heart of the law can deepen this conversation, improving alignment while promoting democratic pluralism.

\section{Positivist Theories for Pluralistic AI Alignment}

The law has long confronted the same problems of pluralism and specification that alignment is now facing, and legal scholars have developed a variety of theories for how it should respond to them \citep{Hart, Holmes, Bickel, Dworkin}. Modern American law makes rules at various levels and takes into its scope much of life and the varieties of human interaction, abstracting norms and patterns of behavior \cite{Lessig}. But each law matters in how it is applied in concrete cases involving the rights and obligations of people, and in fact has no real meaning until such applications--advisory opinions are in fact banned in American federal courts to avoid groundless rulemaking \cite{Kannan}. The question then is how to ensure that general rules bind into the future in a way that protects rights and is predictable such that people can live their lives knowing how the law affects them. Two legal theories explored below suggest that cases, applications of a law over time, can fill in the meaning of general law in a way that allows for specification while preserving space for pluralist disagreement over values.

\subsection{Specification Through Rules and Cases}

Specifying the meaning of legal rules in a predictable and reasonable way is essential to the rule of law and to avoiding arbitrariness and abuses of power \cite{Bickel, Holmes}. But natural language is ambiguous and contextual. Sometimes bees are legally fish \cite{Sanders}. Professor H.L.A. Hart, likely still the leading scholar of jurisprudence \cite{Leiter}, sought to provide an account of adjudication that showed how the meaning of rules is determined in the face of what Hart called this “open texture” of language \cite{Hart}.

The problem that Hart saw was that ambiguity in law persisted despite attempts to add definitions and explanations. Hart’s famous example \cite{Hart2} is a sample rule, perhaps made by a town council, that “forbids you to take a vehicle into the public park.” At first blush, this rule seems reasonable and easy to understand: it disallows cars and motorcycles from entering the park. But complications arise for cases like bicycles, roller skates, and baby strollers that seem to clearly fit the definition of vehicle but also seem unreasonable to exclude from the park \cite{Sargentich}. And what about a military truck intended to be mounted as a memorial \cite{Fuller}? An ambulance coming to the aid of a heart attack victim? Further clarification of the rule is necessary to resolve these cases, and the lengthy law codes risk expanding infinitely more with definitions and exceptions, each with their own definitions and exceptions. This specification problem is of course much worse as soon as the rule at issue moves away from concrete things like vehicles and toward abstractions like justice, helpfulness, bias, or pluralism.

Hart’s solution \cite{Hart} is to say that in language there is a settled core of meaning and a penumbra, at which the edge cases are disputed. When an interpreter of a legal rule is at the core, they should just apply the words at issue as the settled core dictates: no cars in the park. When they are at the fringes, faced with a truly indeterminate case, they must exercise discretion to decide how to interpret the language at issue in light of the social context, underlying goals, or other external features of the situation that might provide guidance. Perhaps there’s evidence that the council wanted to avoid accidents from high speeds: then cars are out and bikes might be too, but tricycles are in \citep{Hart, Dworkin}.

But importantly, Hart’s system has precedents, past decisions and interpretations that indicate how a new dispute on the same issue should be resolved, and these precedents allow the bootstrapping of the theory of open texture into a relatively complete system of specification that retains a degree of flexibility and contextuality. A case is decided and a precedent interpreting some phrase made. Another dispute about the phrase arises and another decision is made, this time with reference to the first one. Over time, a kind of constellation of precedent arises around the phrase at issue that allows for definition of the permissible interpretations of the language at issue by means of triangulation among the precedents. The map of meaning is filled out and new cases can relatively easily be fit into the framework of existing law. The original rule at issue gets content through the process of application and people can predict that later interpretations will resemble earlier ones. Thus, specification of meaning is achieved through the interplay of general rules and concrete applications and a law is at once comprehensible and predictable and also flexible enough that it can be applied appropriately in truly novel cases when the need arises \cite{Hart}.

\subsection{Analogy and Incompletely Theorized Agreements}

Professor Cass Sunstein took the argument a step further, claiming that in fact the law has little need of general principles but rather consists only of analogical reasoning among concrete cases \cite{Sunstein1}, and that this rejection of general principles gives it a special power to protect pluralism \cite{Sunstein2}. For Sunstein, the engine of the law is analogical reasoning:\footnote{One study has shown that analogical reasoning emerges in large language models \cite{webb2023emergentanalogicalreasoninglarge}.} “fixed points” of precedent are established through judicial decisions and then new cases are decided through incremental analogy to these fixed points \citep{Sunstein1, SunsteinAI}.\footnote{Feng et al. \cite{feng2023caserepositoriescasebasedreasoning} cite Sunstein in their case-based reasoning work. However, the approach here differs from their approach, which focuses on modifications of seed cases by experts rather than using cases to democratically give meaning to rules.} Importantly, the justification for the decisions is less relevant than the decisions themselves. Such a low-level, specific approach has important benefits for pluralism. Because the law, at least in adjudication, focuses on concrete cases, there is no need for debates over high level concepts of the good or justice about which people disagree. Instead, two people might agree that a given decision in the case at issue is the right one even if their philosophical justifications for that decision are incompatible \cite{Sunstein2}. For Sunstein, such a mode of reasoning is necessary for a pluralistic society because it allows for cohabitation among possibly incommensurable values. Unlike Hart, who believed that rules are the core of the law and cases only subsidiary ways of informing their meaning, Sunstein argues that it is the cases that actually function as the core, and that the announcement of general principles risks stamping out differing views.

Alignment on particulars without consensus on principles creates what Sunstein \cite{Sunstein2} calls “incompletely theorized agreements,” which allow for the operation of law in society subject to democratic control and protective of plural perspectives. Incompletely theorized agreements are practical and concrete and provide the substance of the “fixed points,” discussed above, from which analogical reasoning to bring new cases within the law can occur. While Sunstein acknowledges that certain higher-level agreements are possible and indeed necessary both on some substantive questions and as “secondary” rules that provide the rules of the game that is law \citep{SunsteinAI, Hart}, emphasizing low-level decisions on concrete questions allows for people with diverse views to live together in society. Combining these pictures, we get a vision of law in which people can come to agreement on high-level questions in and underlying the law where possible but still retain the ability to cooperate where agreement is not. Specification is improved and the ability to move across levels of agreement while still creating law to address novel problems ensures that pluralism is protected.

\section{Discussion}

Alignment must represent the values of the society from which AI emerges, but existing approaches to incorporating democracy into alignment do not overcome either the specification or the pluralism problem. Legal theory can help. Approaches like Anthropic’s Collective Constitution \cite{GanguliCCAI}, powered by reinforcement learning from AI feedback, currently cannot specify how a model should concretely interpret the rules and principles that have been put into it; instead they “leave[] the interpretation of [the principles] to AI systems themselves” based on the models’ pretraining \cite{kundu2023specificversusgeneralprinciples}. “Good for humanity,” one example of a general principle, obviously has this problem, but, as Hart demonstrated, so do all other words in rules. Simply having a group of people deliberate over what principles should be put in the model does not solve this problem and will likely lead instead to vague agreement over good-sounding concepts without a clear sense of what they mean. These existing democratic approaches similarly struggle to preserve pluralistic values. Returning to the Collective Constitution as a leading example, even where the population that performs the deliberation is representatively selected \cite{Flanigan}, the principles selected to be put into models are chosen based on the majority vote of those chosen to provide democratic input, despite substantial minorities of the voters disagreeing with the choices of the majority \citep{GanguliCCAI, fleisig2024majoritywrongmodelingannotator}. This majoritarian selection reinforces the extent to which large language models can be understood as majoritarian machines, making completions based on the predominant perspectives in their training data and finetuning.

The legal theories discussed above can improve how democratic deliberation on alignment is done. Processes like the Collective Constitution \cite{GanguliCCAI} should be supplemented by the inclusion of cases and deliberation about principles of analogy to improve the degree of specification and pluralism that they allow. In one implementation of this approach, instead of merely having users write, deliberate about, and vote on abstract statements of principle, when a user selects a given statement of principle as one that they might agree with, an LLM could be used to generate a set of examples of the application of that statement in different circumstances. Users would then indicate which of the applications they feel matches their understanding of the statement, and, channeling Hart, these applications could be used to better specify how the user actually thinks about the statement that they have selected. A concept like “fairness” that is notoriously hard to define for humans, let alone computers \cite{dwork2011fairnessawareness, dwork2023pseudorandomnessmultigroupfairness}, becomes more tractable by leveraging contextual linguistic reasoning through cases. Next, drawing on Sunstein \citep{Sunstein1, Sunstein2, SunsteinAI}, the cases that users reasoned about could be compared to each other and similar decisions picked out. Data on users who selected different statements of principle but agreed to the same or similar decision of analogous concrete cases could help alignment researchers understand the underlying similarities across plural perspectives that make people decide the same way for different reasons and allow researchers to incorporate incompletely theorized agreements into models’ reasoning processes. Ideally, the result of this process would be improved measures of alignment across a variety of cases. But it should also allow for more effective intercultural and global reasoning about how to align AI as well, increasing the extent to which models are pluralist.

The above proposal is merely a start, and much more work is needed to turn the idea into an implementable alignment technique. In particular, it is unclear how to get the models used to produce concrete cases to cover the whole spectrum of possible meanings of the word that the cases are intended to illustrate, especially if they are by their nature more likely to produce outputs near the most likely or core meaning of a word. Even if the approach were to be implemented and positive results demonstrated from the experiments, there would likely remain difficulties in indicating to a model how to resolve problems that actually involved disputes over higher values and rules rather than specific cases, which some leading legal theorists believe to be the core of the law \cite{Dworkin}. Questions of structure and hierarchy that operate at the core of constitutional law would have to be resolved. Other secondary rules \cite{Hart}, including the extent to which different decisions became precedent that weighed against others, would have to be decided, and encouraging people to reason about those issues might be harder than thinking about direct issues of value and principle about which people tend to have intuitions. Ultimately, the best that the law has done is provide directions \cite{SunsteinAI} and "rules of thumb" \cite{Dworkin}, and legal theory likely has as much to learn from alignment as vice versa. Nonetheless, this paper shows that legal theory is a rich vein upon which alignment should begin to draw and suggests both practical and theoretical avenues to do so.

\section{Conclusion}

AI alignment faces two key problems, pluralism and specification, that legal theory can help solve. The law has long faced the difficulty of how to have general rules that can be predictably applied in specific situations and that avoid stamping out the forms of life of those who have other values and visions of the good. Two leading legal theories suggest that the law solves these problems through the use of concrete cases that fill in the meaning of general rules in their application and that provide space for disagreement about values but alignment on outcomes. The map of possible meanings of a given term is filled out over time in a way that remains close to the actual meaning of practical cases, and people can agree on implementation while retaining different views on principle. These insights should be used to supplement existing attempts to put democratic deliberation into AI, and I have sketched out one potential approach to doing so that makes use of large language models and their capabilities to generate cases to illustrate alignment principles. However, empirical testing remains to be done. Collaboration between law and AI can help ensure that AI is aligned to diverse human values and that it becomes a force that helps people live their lives in the way best for them.

\newpage
\bibliographystyle{plain}
\bibliography{LTAI.bib}

\end{document}